\documentstyle[12pt]{article}
\begin{document}
\newlength{\lno}\lno1.5cm
\newlength{\len}\len=\textwidth\addtolength{\len}{-\lno}

\baselineskip6mm
\newcommand{\be}{\begin{equation}\label}\newcommand{\ee}{\end{equation}}
\newcommand{\bea}{\begin{eqnarray*}}\newcommand{\eea}{\end{eqnarray*}}
\def\eqa#1#2{\\\parbox{\len}{\bea#2\eea}\hfill\parbox{\lno}{\be{#1}\ee}}
\newcommand{\ban}{\begin{eqnarray}\label}\newcommand{\ean}{\end{eqnarray}}
\def\nn{\nonumber}
\newcommand{\bp}{\begin{picture}}\newcommand{\ep}{\end{picture}}
\newcommand{\ba}{\begin{array}}\newcommand{\ea}{\end{array}}
\newcommand{\s}{\scriptstyle}
\newcommand{\x}{{\underline x}}
\newcommand{\ua}{{\underline\alpha}}
\newcommand{\ub}{{\underline\beta}}
\newcommand{\uu}{{\underline u}}
\renewcommand{\theequation}{\mbox{\arabic{section}.\arabic{equation}}}
\newcommand{\la}{\langle\,}\newcommand{\ra}{\,\rangle}
\newcommand{\mi}{\,|\,}

\newtheorem{theo}{Theorem}[section]
\newtheorem{cond}[theo]{Conditions}
\newtheorem{defi}[theo]{Definition}
\newtheorem{exam}[theo]{Example}\newtheorem{lemm}[theo]{Lemma}
\newtheorem{rema}[theo]{Remark}
\newtheorem{prop}[theo]{Proposition}
\newtheorem{coro}[theo]{Corollary}
\newtheorem{beth}[theo]{BETHE ANSATZ}

\def\V{{V^{1\dots n}}}
\def\O{{\Omega^{1\dots n}}}
\def\f{{f^{1\dots n}}}

\def\A{{A_{1\dots n}}}
\def\C{{C_{1\dots n}}}
\def\D{{D_{1\dots n}}}
\def\einsop{{\bf 1}}

\def\V{{V^{1\dots n}}}
\def\VV{{{V^{(1)}}^{\bar1\dots\bar n_-b_1\dots b_m}}}
\def\ff{{{f^{(1)}}^{\bar1\dots \bar n_-b_1\dots b_m}}}
\def\g#1{{g^{1\dots n#1}}}
\def\gg{{g^{1\dots nb_1\dots b_m}}}
\def\T#1{{T_{1\dots n,#1}}}
\def\B#1{{B_{1\dots n,#1}}}

\def\Tq#1{{T_{1\dots n,#1}^Q}}
\def\Ttq#1{{\tilde T_{\bar #1,1\dots n}^Q}}

\title{$U(N)$ Matrix Difference Equations\\ and a Nested Bethe Ansatz}
\author{\\
H. Babujian$^{1,2,3}$, M. Karowski$^4$ and A. Zapletal$^{5,6}$
\\{\small\it Institut f\"ur Theoretische Physik}\\
{\small\it Freie Universit\"at Berlin, Arnimallee 14, 14195 Berlin, Germany} 
}
\date{\small\today}

\maketitle
\footnotetext[1]{Permanent address: Yerevan Physics Institute,
Alikhanian Brothers 2, Yerevan, 375036 Armenia.}
\footnotetext[2]{Partially supported by the grant 211-5291 YPI of the German
Bundesministerium f\"ur Forschung und Technologie and
through the VW project 'Cooperation with scientists from CIS'.}
\footnotetext[3]{e-mail: babujian@lx2.yerphi.am}
\footnotetext[4]{e-mail: karowski@physik.fu-berlin.de}
\footnotetext[5]{Supported by DFG, Sonderforschungsbereich 288
'Differentialgeometrie und Quantenphysik'}
\footnotetext[6]{e-mail: zapletal@physik.fu-berlin.de} 

\begin{abstract}
A system of $U(N)$-matrix difference equations is solved by means of a
nested version of a generalized Bethe Ansatz. The highest weight property of
the solutions is proved and some examples of solutions are calculated
explicitly.
\end{abstract}
\section{Introduction}
This is the second part of a series of articles on the generalized nested
Bethe Ansatz to solve matrix difference equations for different symmetry
groups. The first part \cite{BKZ} (later denoted by [I]) was on the $SU(N)$
problem. In this part we investigate $U(N)$ symmetric matrix difference
equations which are considerable more complicated than in the $SU(N)$
case, because the Bethe Ansatz reference states are more complicated.
The reader is assumed to be familiar with the techniques
and results of part [I]. We concentrate here on the problems which have
to be solved in addition to those of part [I].
The results of this paper will be used in \cite{BFK}
to calculate exact form factors for the chiral Gross-Neveu model.

The article is organized similar as [I].
In Section~\ref{s2} we recall some results concerning the
$U(N)$ R-matrix.
In Section~\ref{s3} we introduce the nested generalized Bethe Ansatz to
solve a system of $U(N)$ difference equations and present the solutions
again in terms of "Jackson-type Integrals". In addition to the one used in
[I] we have to introduce an additional monodromy matrix
of the new type also fulfilling a new type of Yang-Baxter relation.
In Section~\ref{s4} we prove the highest weight property of the solutions
and calculate the weights.
Section~\ref{s5} contains some examples of solutions.
\setcounter{section}{1}\setcounter{equation}{0}
\section{The $U(N)$  R - matrix}\label{s2}
Let $\V$ be the tensor product space
\be{2.1}\V=V_1\otimes\dots\otimes V_n\ee
where the vector spaces are $V_i=(V\oplus\bar V)_i$. Here
$V\cong{\bf C^N}$ is considered as a fundamental representation space of
$U(N)$ and $\bar V\cong{\bf C^N}$ as the conjugate representation space.
The vectors of $V$ are sometimes called "particles" with positive $U(1)$
charge and those of $\bar V$ "anti-particles" with negative  $U(1)$
charge. We will also use instead of $1,2,\dots$ the indices $a,b$, etc. 
to refer to the spaces $V_a,V_b$, etc.
It is straightforward
to generalize the results of this paper to the case where the $V_i$
are vector spaces for other representations.
We denote the canonical basis vectors by
\be{2.2}
\mi\alpha_1\dots\alpha_n\ra\in
\V,~~\Big(\alpha_i=(1,+),\dots,(N,+),(1,-),\dots,(N,-)\Big).
\ee
We will later also use the simpler notation $(\alpha,+)\equiv\alpha$ for
particles and $(\alpha,-)\equiv\bar\alpha$ for anti-particles.
A vector in $\V$ is denoted by  $v^{1\dots n}$ and a matrix acting in $\V$
by $\A\in End(\V)$.

The $U(N)$ spectral parameter depending R-matrix \cite{BKKW} acting on the
tensor product of two particle spaces $V^{12}=V\otimes V$
coincides with that of $SU(N)$ used in [I].
It may be written and depicted as
\be{2.3}
R_{12}(x_1-x_2)=b(x_1-x_2)\einsop_{12}+c(x_1-x_2)\,P_{12}~~=~~
\ba{c}
\unitlength2.7mm
\bp(5,4)
\put(1,0){\line(1,1){4}}
\put(5,0){\line(-1,1){4}}
\put(1,0){\vector(1,1){1.3}}
\put(5,0){\vector(-1,1){1.3}}
\put(3,2){\vector(1,1){1.3}}
\put(3,2){\vector(-1,1){1.3}}
\put(0,.5){$\s x_1$}
\put(5,.5){$\s x_2$}
\ep
\ea
~~~~:\,V^{12}\to V^{21},
\ee
where $P_{12}$ is the permutation operator.
For $U(N)$ there is in addition the R-matrix acting on the tensor product of
the anti-particle particle spaces $V^{\bar12}=\bar V\otimes V$
\be{2.4}
R_{\bar12}(x_1-x_2)=\einsop_{\bar12}+d(x_1-x_2)\,K_{\bar12}~~=~~
\ba{c}
\unitlength2.7mm
\bp(5,4)
\put(1,0){\line(1,1){4}}
\put(5,0){\line(-1,1){4}}
\put(3,2){\vector(-1,-1){1.3}}
\put(5,0){\vector(-1,1){1.3}}
\put(5,4){\vector(-1,-1){1.3}}
\put(3,2){\vector(-1,1){1.3}}
\put(0,.5){$\s x_1$}
\put(5,.5){$\s x_2$}
\ep
\ea
~~~~:\,V^{\bar12}\to V^{2\bar1},
\ee
where $K_{\bar12}$ is the annihilation-creation matrix.
There is no particle-anti-particle "backward scattering"
(see e.g. \cite{BKKW}).
Here and in the following we associate a variable (spectral parameter)
$x_i\in{\bf C}$ to each space $V_i$ which is graphically represented by a
line labeled by $x_i$ (or simply by $i$). In addition an arrow on the lines
denotes the $U(1)$-charge flow.
(If we do not want to specify the direction of charge flow, we draw no
arrow.)
The components of the R-matrices are
\be{2.5}
R_{\alpha\beta}^{\delta\gamma}=
\delta_{\alpha\gamma}\delta_{\beta\delta}\,b+
\delta_{\alpha\delta}\delta_{\beta\gamma}\,c~~=~
\ba{c}
\unitlength2mm
\bp(6,6)
\put(1,1){\line(1,1){4}}
\put(5,1){\line(-1,1){4}}
\put(1,1){\vector(1,1){1.3}}
\put(5,1){\vector(-1,1){1.3}}
\put(3,3){\vector(1,1){1.3}}
\put(3,3){\vector(-1,1){1.3}}
\put(.5,0){$\s\alpha$}
\put(5,0){$\s\beta$}
\put(5,5.3){$\s\gamma$}
\put(.5,5.3){$\s\delta$}
\ep
\ea~,~~
R_{\bar\alpha\beta}^{\delta\bar\gamma}=
\delta_{\alpha\gamma}\delta_{\beta\delta}+
\delta_{\alpha\beta}\delta_{\delta\gamma}\,d~~=~
\ba{c}
\unitlength2mm
\bp(6,6)
\put(1,1){\line(1,1){4}}
\put(5,1){\line(-1,1){4}}
\put(3,3){\vector(-1,-1){1.3}}
\put(5,1){\vector(-1,1){1.3}}
\put(5,5){\vector(-1,-1){1.3}}
\put(3,3){\vector(-1,1){1.3}}
\put(.5,0){$\s\bar\alpha$}
\put(5,0){$\s\beta$}
\put(5,5.3){$\s\bar\gamma$}
\put(.5,5.3){$\s\delta$}
\ep
\ea
\ee
with $\alpha=(\alpha,+)$ etc. and $\bar\alpha=(\alpha,-)$ etc.
The functions
\be{2.6}
b(x)=\frac x{x-2/N},~~c(x)=\frac{-2/N}{x-2/N},~~d(x)=\frac{2/N}{x-1}.
\ee
belong to the rational solution of the Yang-Baxter equations \cite{BKKW}
$$R_{12}(x_1-x_2)\,R_{13}(x_1-x_3)\,R_{23}(x_2-x_3)
=R_{23}(x_2-x_3)\,R_{13}(x_1-x_3)\,R_{12}(x_1-x_2).
$$$$
\unitlength4mm
\bp(9,4)
\put(0,1){\line(1,1){3}}
\put(0,3){\line(1,-1){3}}
\put(2,0){\line(0,1){4}}
\put(4.3,2){$=$}
\put(6,0){\line(1,1){3}}
\put(6,4){\line(1,-1){3}}
\put(7,0){\line(0,1){4}}
\put(.2,.5){$\s 1$}
\put(1.3,0){$\s 2$}
\put(3,.2){$\s 3$}
\put(5.5,.2){$\s 1$}
\put(7.2,0){$\s 2$}
\put(8.4,.4){$\s 3$}
\ep~~
$$
which holds for all possible charge flows.
The matrices $R_{1\bar2}$ and $R_{\bar1\bar2}$ have the same matrix elements
as $R_{\bar12}$ and $R_{12}$, respectively. The inversion ("unitarity")
relation of the R-matrix reads and may be depicted as
\be{2.7}
R_{21}(x_2-x_1)\,R_{12}(x_1-x_2)=1~:~~~~~
\ba{c}
\unitlength3mm
\bp(9,4)
\put(1,0){\line(1,1){2}}
\put(3,0){\line(-1,1){2}}
\put(1,2){\line(1,1){2}}
\put(3,2){\line(-1,1){2}}
\put(7,0){\line(0,1){4}}
\put(9,0){\line(0,1){4}}
\put(4.5,1.7){$=$}
\put(.2,0){$\s 1$}
\put(3.2,0){$\s 2$}
\put(6.2,0){$\s 1$}
\put(8.2,0){$\s 2$}
\ep
\ea
\ee
again for all possibilities charge flows.
A further property of the $U(N)$ R-matrices is crossing and may be
written and depicted as
$$
R_{\alpha\beta}^{\delta\gamma}(x_1-x_2)
=b(x_1-x_2)\,R_{\bar\delta\alpha}^{\gamma\bar\beta}((x_2+1)-x_1)
=b(x_1-x_2)\,R_{\beta\bar\gamma}^{\bar\alpha\delta}(x_2-(x_1-1))
$$$$
\ba{c}
\unitlength3mm
\bp(4,6)
\put(0,1){\line(1,1){4}}
\put(4,1){\line(-1,1){4}}
\put(0,1){\vector(1,1){1.3}}
\put(4,1){\vector(-1,1){1.3}}
\put(2,3){\vector(1,1){1.3}}
\put(2,3){\vector(-1,1){1.3}}
\put(0,0){$\s\alpha$}
\put(3.7,0){$\s\beta$}
\put(0,5.5){$\s\delta$}
\put(3.7,5.5){$\s\gamma$}
\ep
\ea
=b~~~
\ba{c}
\unitlength3mm
\bp(6,6)
\put(1,1){\line(1,1){4}}
\put(4,1){\line(-1,2){2}}
\put(3,3){\vector(-1,-1){1.2}}
\put(4,1){\vector(-1,2){.6}}
\put(5,5){\vector(-1,-1){1.2}}
\put(3,3){\vector(-1,2){.6}}
\put(1,5){\oval(2,8)[lb]}
\put(5,1){\oval(2,8)[tr]}
\put(3.5,0){$\s\alpha$}
\put(5.7,0){$\s\beta$}
\put(0,5.5){$\s\delta$}
\put(2,5.5){$\s\gamma$}
\ep
\ea
=b~~~
\ba{c}
\unitlength3mm
\bp(6,6)
\put(2,1){\line(1,2){2}}
\put(5,1){\line(-1,1){4}}
\put(2,1){\vector(1,2){.6}}
\put(3,3){\vector(1,-1){1.2}}
\put(3,3){\vector(1,2){.6}}
\put(1,5){\vector(1,-1){1.2}}
\put(1,1){\oval(2,8)[lt]}
\put(5,5){\oval(2,8)[br]}
\put(0,0){$\s\alpha$}
\put(2,0){$\s\beta$}
\put(4,5.5){$\s\delta$}
\put(6,5.5){$\s\gamma$}
\ep
\ea
$$
where again we have used the notation $\alpha=(\alpha,+)$ etc. and
$\bar\alpha=(\alpha,-)$ etc.
We have introduced the graphical rule that a line changing the
"time direction" also interchanges particles and anti-particles and changes
$x\to x\pm 1$ as follows
\be{2.8}
\ba{c}
\unitlength4mm
\bp(14,3)
\put(2,1){\oval(2,4)[t]}
\put(0,1){$\s x$}
\put(3.3,1){$\s x-1$}
\put(-.2,0){$\s(\alpha,\pm)$}
\put(2.2,0){$\s(\alpha,\mp)$}
\put(11,2){\oval(2,4)[b]}
\put(9,1){$\s x$}
\put(12.3,1){$\s x+1$}
\put(8.8,2.5){$\s(\alpha,\pm)$}
\put(11.2,2.5){$\s(\alpha,\mp)$}
\ep~.
\ea
\ee
We introduce a monodromy matrix (with $\x=x_1,\dots,x_n$)
\be{2.9}\T{0}(\x;x_0)=R_{10}(x_1-x_0)\dots R_{n0}(x_n-x_0)\ee
$$
=
\ba{c}
\unitlength2.8mm
\bp(14,4)
\put(0,2){\line(1,0){14}}
\put(14,2){\vector(-1,0){1}}
\put(2,0){\line(0,1){4}}
\put(6,0){\line(0,1){4}}
\put(6,2){\vector(0,-1){1.5}}
\put(6,4){\vector(0,-1){1.5}}
\put(8,0){\line(0,1){4}}
\put(8,0){\vector(0,1){1}}
\put(8,2){\vector(0,1){1}}
\put(12,0){\line(0,1){4}}
\put(1,0){$\s1$}
\put(5,0){$\s i$}
\put(7,0){$\s j$}
\put(11,0){$\s n$}
\put(13,1){$\s 0$}
\put(3,3){$\dots$}
\put(9,3){$\dots$}
\put(6.8,3){$.$}
\ep
\ea
$$
as a matrix acting in the tensor product of the "quantum space" $\V$ and
the "auxiliary space" $V_0\cong V\cong{\bf C}^N$ as a particle space.
Since there is no "charge reflection" the positions of the particles
and the anti-particles will not change under the application of $\T{0}$ to a
state in $\V$.
(The construction of the Bethe states will be not symmetric with respect
to particles and anti-particles, because we only use the monodromy matrix
with particles for the auxiliary space.)

The Yang-Baxter algebra relations
\be{2.10}
\T{a}(\x;x_a)\,\T{b}(\x;x_b)\,R_{ab}(x_a-x_b)
=R_{ab}(x_a-x_b)\,\T{b}(\x;x_b)\,\T{a}(\x;x_a)
\ee
imply the basic algebraic properties of the sub-matrices w.r.t the
auxiliary space defined by
\be{2.11}
{T_{1\dots n}}^\alpha_\beta(\x;x)\equiv
\left(\matrix{\A(\x;x)&\B{\beta}(\x;x)\cr
\C^\alpha(\x;x)&\D^\alpha_\beta(\x;x)}\right).
\ee
The indices $\alpha,\beta$ on the left hand side run from 1 to $N$ and on the
right hand side from 2 to $N$.
The commutation rules of the $A,B,C,D$ are the same as for the $SU(N)$ case
of [I].

\setcounter{section}{2}\setcounter{equation}{0}
\section{The $U(N)$ - difference equation}\label{s3}
Let 
$$\f(\x)=~~
\ba{c}
\unitlength4mm
\bp(7,4)
\put(3.5,1){\oval(7,2)}
\put(3.5,1){\makebox(0,0){$f$}}
\put(1,2){\line(0,1){2}}
\put(3,2){\line(0,1){2}}
\put(3,4){\vector(0,-1){1.5}}
\put(4,2){\line(0,1){2}}
\put(4,2){\vector(0,1){1}}
\put(6,2){\line(0,1){2}}
\put(0,3.5){$\s x_1$}
\put(2,3.5){$\s x_i$}
\put(1.4,3){$\dots$}
\put(3.4,3){$.$}
\put(4.2,3.5){$\s x_j$}
\put(6.2,3.5){$\s x_n$}
\put(4.4,3){$\dots$}
\ep
\ea
~~\in\V
$$
be a vector valued function of $\x=x_1,\dots,x_n$ with values in $\V$.

Similar as in [I] we constrain the function $\f$ by the 
\begin{cond}\label{cond}
The following symmetry and periodicity conditions on the vector
valued function $\f(\x)$ are supposed to be valid:
\begin{itemize}
\item[\rm(i)] The symmetry property under the exchange of two neighboring
spaces $V_i$ and $V_j$ and the variables $x_i$ and $x_j$, at the same time,
is given by
\be{3.1}
f^{\dots ji\dots}(\dots,x_j,x_i,\dots)=
R_{ij}(x_i-x_j)\,f^{\dots ij\dots}(\dots,x_i,x_j,\dots).
\ee
\item[\rm(ii)]
The {\bf system of matrix difference equations} holds
\be{3.2}
\fbox{\rule[-3mm]{0cm}{8mm} $
\f(\dots,x_i+2,\dots)=Q_{1\dots n}(\x;i)\,\f(\dots,x_i,\dots)
~,~~(i=1,\dots,n)$ }
\ee
where the matrices $Q_{1\dots n}(\x;i)\in End(\V)$ are defined by
\be{3.3}
Q_{1\dots n}(\x;i)=R_{i+1i}(x_{i+1}-x_i')\dots R_{ni}(x_n-x_i')\,
R_{1i}(x_1-x_i)\dots R_{i-1i}(x_{i-1}-x_i)
\ee
with $x'_i=x_i+2$.
\end{itemize}
\end{cond}
The Yang-Baxter equations for the R-matrix guarantee that these conditions
are compatible. 
The shift of $2$ in eq.~(\ref{3.2}) could be replaced by
an arbitrary $\kappa$ for the application to the form factor
problem, however, it is fixed to $2$ because of crossing symmetry.
The $Q_{1\dots n}(\x;i)$ fulfill the same commutation rules as in [I].

Proposition 3.2 and Remark 3.3 of [I] also hold for the $U(N)$ case.
Also we need the new type of monodromy matrix given by Definition 3.4 of [I]
with
\be{3.4}
Q_{1\dots n}(\x;i)={\rm tr}_0\,\Tq{0}(\x;i).
\ee
We use this formula if the index $i$ is associated to a particle.
If the index $i$ is associated to an anti-particle,
we will use a further monodromy matrix of a new type
given by the following
\begin{defi} 
For $i=1,\dots,n$
\be{3.5}
\Ttq{0}(\x;i)=R_{\bar 01}(x_i-x_1)\dots R_{\bar0i-1}
(x_i-x_{i-1})\,
P_{\bar0\bar i}\,R_{\bar0i+1}(x'_i-x_{i+1})\dots R_{\bar0n}(x'_i-x_n)
\ee
with the auxiliary space $V_{\bar0}=\bar V$ and $x'_i=x_i+2$.
\end{defi} 
By the "unitarity" of the R-matrix (\ref{2.7}) the trace of this monodromy
matrix is the inverse of the Q-matrix
\be{3.6}
Q^{-1}_{1\dots n}(\x;i)={\rm tr}_{\bar0}\,\Ttq{0}(\x;i).
\ee
The new type of monodromy matrix $\Ttq{0}$ fulfills a new type of
Yang-Baxter relation in the form
\be{3.7}
\Ttq{a}(\x;i)\,R_{\bar ab}(x_i-u)\,\T{b}(\x',u)
=\T{b}(\x,u)\,R_{\bar ab}(x'_i-u)\,\Ttq{a}(\x;i)
\ee
with $\x'=x_1,\dots,x'_i,\dots,x_n$ and $x'_i=x_i+2$.
In addition to the commutations rules (3.10) and (3.11) of [I] for the case
that the indices $i$ and $a$ belong to particles, we have the commutation
rules for the case that the indices $i$ and $a$ correspond to anti-particles
\be{3.8}
\tilde A^Q(\x;i)\,B_b(\x',u)=\frac{1-d^2(x_i'-u)}{1+d(x_i-u)}\,
B_b(\x',u)\,\tilde A^Q(\x;i)
+d(x_i'-u)\,\tilde B^Q_{\bar b}(\x;i)\,A(\x',u)
\ee
\be{3.9}
\tilde D^Q_{\bar a}(\x;i)\,B_b(\x',u)
=B_b(\x,u)\,R_{\bar ab}(x'_i-u)\,\tilde D^Q_{\bar a}(\x;i)
-d(x_i-u)\,\tilde B^Q_{\bar a}(\x;i)\,K_{\bar ab}\,D_b(\x',u).
\ee
The system of difference equations (\ref{3.2}) can be solved by means of a
generalized ("off-shell") nested Bethe Ansatz.
The first level is given by the
\begin{beth}\label{be}
\be{3.10}
\f(\x)=\sum_\uu~\B{\beta_m}(\x,u_m)\dots\B{\beta_1}(\x,u_1)\,
\g{\beta_1\dots\beta_m}(\x,\uu)
\ee
where summation over $1<\beta_1,\dots,\beta_m\le N$ is assumed.
The summation over $\uu$ is specified by
\be{3.11}
\uu=(u_1,\dots,u_m)=(\tilde u_1-2l_1,\dots,\tilde u_m-2l_m)
~,~~~l_i\in{\bf Z},
\ee
where the $\tilde u_i$ are arbitrary constants.
For a fixed set of indices $\ub=\beta_1,\dots,\beta_m$ $(1<\beta_i\le N)$
the reference state vector $\g{\ub}\in \V$ fulfills
\be{3.12}
\C^\gamma\,\g{\ub}=0~~(1<\gamma\le N)
\ee
\end{beth}
Note that here we have denoted by $g$, what was called $\Omega\,g$ in [I].
This notation is more convenient, since contrasting the $SU(N)$ case, the
space of reference states is here higher dimensional for $U(N)$.
This space $V_{ref}$ is spanned by all basis vectors of the form
$$
\mi\alpha_1,\dots,\alpha_n\ra~,~~\alpha_i=\cases{(1,+)& for particles\cr
(\alpha_i,-)~,~~(1<\alpha_i\le N)~& for antiparticles.}
$$
The $U(1)$-charge is left invariant by operators like the monodromy matrices
$\T{0}$, $\Tq{0}$, $\Ttq{0}$ and also by the operations of Conditions
\ref{cond}.
Therefore the space $\V$ decomposes into invariant subspaces of fixed
charge, i.e. fixed numbers
$n_-,n_+~(n_-+n_+=n)$ of anti-particles and particles, respectively.
Moreover because of the symmetry property (i) it is sufficient to consider
the even smaller subspaces
where the anti-particles and the particles are sitting at fixed places
\be{3.13}
V^{1\dots n}_{n_-}=\bigoplus_{I_-\subset I}V^{1\dots n}_{I_-}~,~~
(I_-\cup I_+=I=\{1,\dots,n\},~|I_-|=n_-,~|I_+|=n_+).
\ee
Here $I_-$ and $I_+$ denote the positions of the anti-particles and
particles, respectively.
The space of reference states in the subspace $V^{1\dots n}_{I_-}$ may be
written as
\be{3.14}
V_{ref}\cap V^{1\dots n}_{I_-}\cong
\bar V^{(1)}_1\otimes\dots\otimes\bar V^{(1)}_{n_-}\otimes
\Omega^{1\dots n_+}
\ee
where the $\bar V^{(1)}_i\cong{\bf C}^{N-1}$
contain only anti-particle states with $\alpha>1$ and with the fixed vector
$$
\Omega^{1\dots n_+}=\mi1\dots1\ra\in\bigotimes_{i\in I_+}V_i
$$
Therefore $g$ may also be considered as a vector
$$
\gg\in\VV\otimes\Omega^{1\dots n_+}
$$
with
\be{3.15}
\VV=\bigotimes_{i\in I_-}\bar V^{(1)}_i\otimes
V^{(1)}_{b_1}\otimes\dots\otimes V^{(1)}_{b_m}
\ee
where the $V^{(1)}_{b_i}\cong{\bf C}^{N-1}$ are particle spaces.
Again as in [I] we define a new vector valued function $\ff\in\VV$ by
\begin{defi}\label{d3.2}
Let $\g{\ub}(\x,\uu)$ be given in the subspace $V^{1\dots n}_{I_-}\subset\V$
with fixed positions of particles and anti-particles as
\be{3.16}
\gg(\x,\uu)=
\prod_{i\in I_+}\prod_{j=1}^m\psi(x_i-u_j)\,
\prod_{1\le i<j\le m}\tau(u_i-u_j)\,\ff(\x^{(-)},\uu)\,
\otimes\,\Omega^{1\dots n_+}
\ee
where $\x^{(-)}=\{x_i;i\in I_-\}$ are the spectral parameters of the
anti-particles only.
\end{defi}
The functions $\psi(x)$ and $\tau(x)$ are the same as in [I]
\be{3.17}
\psi(x)=\frac{\Gamma(1-\frac1N+\frac x2)}{\Gamma(1+\frac x2)}
,~~~\tau(x)=\frac{x\,\Gamma(\frac1N+\frac x2)}
{\Gamma(1-\frac1N+\frac x2)}.
\ee
For the case that the anti-particles are sitting at the first $n_-$ places
$I_-=\{1,\dots,n_-\}$ the BETHE ANSATZ \ref{be} may be depicted as
$$
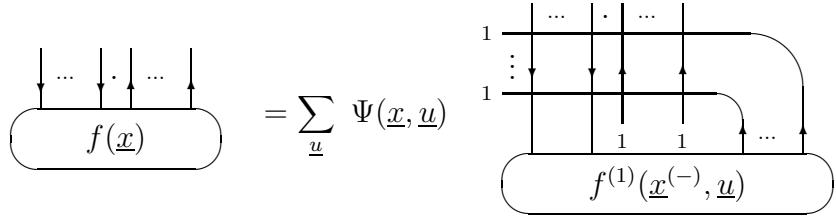
\ba{c}
\unitlength4mm
\bp(7,4)
\put(3.5,1){\oval(7,2)}
\put(3.5,1){\makebox(0,0){$f(\x)$}}
\put(1,2){\line(0,1){2}}
\put(1,4){\vector(0,-1){1.5}}
\put(3,2){\line(0,1){2}}
\put(3,4){\vector(0,-1){1.5}}
\put(4,2){\line(0,1){2}}
\put(4,2){\vector(0,1){1}}
\put(6,2){\line(0,1){2}}
\put(6,2){\vector(0,1){1}}
\put(1.5,3){$\s\dots$}
\put(3.3,3){$.$}
\put(4.5,3){$\s\dots$}
\ep
\ea
~~=\sum_\uu~\Psi(\x,\uu)~
\ba{c}
\unitlength4mm
\bp(12,7)
\put(6.5,1){\oval(11,2)}
\put(6.5,1){\makebox(0,0){$f^{(1)}(\x^{(-)},\uu)$}}
\put(1,3){\oval(16,2)[tr]}
\put(1,3){\oval(20,6)[tr]}
\put(9,2){\vector(0,1){1}}
\put(11,2){\vector(0,1){1}}
\put(2,2){\line(0,1){5}}
\put(2,6){\vector(0,-1){1.5}}
\put(2.5,6.5){$\s \dots$}
\put(4.3,6.5){$.$}
\put(9.5,2.5){$\s \dots$}
\put(1.2,4.5){$\s \vdots$}
\put(5.5,6.5){$\s \dots$}
\put(4,2){\line(0,1){5}}
\put(4,6){\vector(0,-1){1.5}}
\put(5,3){\line(0,1){4}}
\put(5,4){\vector(0,1){1}}
\put(7,3){\line(0,1){4}}
\put(7,4){\vector(0,1){1}}
\put(4.8,2.2){$\s 1$}
\put(6.8,2.2){$\s 1$}
\put(.3,3.8){$\s 1$}
\put(.3,5.8){$\s 1$}
\ep
\ea
$$
where $\Psi(\x,\uu)$ is given by the products of $\psi$'s and $\tau$'s of
eq.~(\ref{3.16}). The main theorem of this article analogous to the one of
[I].
\begin{theo}\label{t3.1}
Let the vector valued function $\f(\x)$ be given by the BETHE ANSATZ
\ref{be} where $g^{1\dots m}(\x,\uu)$ is of the form of Definition
\ref{d3.2}.
If in addition the vector valued function $\ff(\x^{(-)},\uu)\in\VV$ fulfills
the Conditions \ref{cond} {\rm(i)$^{(1)}$ and (ii)$^{(1)}$}, then also
$\f(\x)\in\V$ fulfills the Conditions \ref{cond} {\rm(i) and (ii)}, i.e.
$\f(\x)$ is a solution of the set of difference equations (\ref{3.2}).
\end{theo}
{\bf Proof:} Condition \ref{cond} (i) follows as in [I].
Again it is sufficient to prove Condition \ref{cond} (ii) for $i=n$
$$
Q(\x;n)\,f(\x)=f(\x')~,~~(\x'=x_1,\dots,x'_n=x_n+2).
$$
where the indices $1\dots n$ have been suppressed.
For the case that the index $n$ belongs to a particle the proof is the same
as in [I], with the only difference that here the next level Q-matrix
is given by the trace of
\be{3.17a}
T^{Q(1)}_a(\x^{(-)},\uu;b_m)=\prod_{i\in I_-}R_{ia}(x_i-u_m)\,
R_{b_1a}(u_1-u_m)\dots R_{b_{m-1}a}(u_{m-1}-u_m)\,P_{b_ma}.
\ee

For the case that the index $n$ belongs to an anti-particle
we will prove the inverse of (ii) using $\Ttq{a}$.
We apply $Q^{-1}(\x;n)$ given by the trace of $\Ttq{a}(\x;n)$ to the vector
$\f(\x')$ as given by eq.~(\ref{3.10}) and push $\tilde A^Q(\x;n)$ and
$\tilde D^Q_{\bar a}(\x;n)$
through all the $B$'s using the commutation rules (\ref{3.8}) and
(\ref{3.9}). Again with $\x'=x_1,\dots,x'_n=x_n+2$ we obtain
\bea
\tilde A^Q(\x;n)\,B_{b_m}(\x',u_m)\dots B_{b_1}(\x',u_1)
=B_{b_m}(\x,u_m)\dots B_{b_1}(\x,u_1)~~~~~~\\
\times\prod_{j=1}^m\frac{1-d^2(x_n'-u)}{1+d(x_n-u)}\,
\tilde A^Q(\x;n)+{\rm uw}_A
\eea
\bea
\tilde D^Q_{\bar a}(\x;\bar n)\,B_{b_m}(\x',u_m)\dots B_{b_1}(\x',u_1)
=B_{b_m}(\x,u_m)\dots B_{b_1}(\x,u_1)~~~~~~\\
\times R_{\bar ab_1}(x'_n-u_1)\dots R_{\bar ab_m}(x'_n-u_m)\,
\tilde D^Q_{\bar a}(\x;\bar n)+{\rm uw}_{D_a}
\eea
The "wanted" and "unwanted" terms have origins analogous to the ones
considered in [I].
If we insert these equations into the representation (\ref{3.10}) of
$f(\x')$ we find, however, that the wanted contribution from $A^Q$ vanish
and the wanted contribution from $D^Q$ already gives the desired result.
By definition of $\tilde T^Q$ and the reference state vector $g$ we have
$$\tilde A^Q(\x;n)\,g=0~, ~~~
\tilde D^Q_{\bar a}(\x;n)\,g=
\prod_{i\in I_-}R_{\bar ai}(x_n-x_i)\,P_{\bar a\bar n}\,g$$
if the index $\bar a$ correspond to an anti-particle.
With
$${Q^{(1)}}^{-1}(\x^{(-)},\uu;n)={\rm tr}_a\,\Big(
\prod_{i\in I_-}R_{\bar ai}(x_n-x_i)\,P_{\bar a\bar n}\,
R_{\bar ab_1}(x'_n-u_1)\dots R_{\bar ab_m}(x'_n-u_m)\Big)
$$
and the assumption
$${Q^{(1)}}^{-1}(\x^{(-)},\uu;n)\,f^{(1)}({\x^{(-)}}',\uu)=f^{(1)}(\x^{(-)},\uu)$$
it follows that the wanted term from $\tilde D^Q$ yields $f(\x)$.
Again as in [I] the unwanted contributions can be written as differences
which vanish after summation over the $u$'s.
Because of the commutation relations of the $B$'s (see (2.11) of [I]) and the
symmetry property (i)$^{(1)}$ of $\gg(\x^{(-)},\uu)$ it is sufficient to
consider only the unwanted terms for $j=m$ denoted by uw$_A^m$ and uw$_D^m$.
\bea
{\rm uw}_A^m(\x,\uu)&=&d(x'_n-u_m)\,\tilde B^Q_{\bar b_m}(\x;m)\dots
B_{b_1}(\x,u_1)\prod_{j<m}\frac1{b(u_m-u_j)}\,A(\x,u_m)\\
{\rm uw}_{D_{\bar a}}^m(\x,\uu)&=&-d(x_n-u_m)\,\tilde B^Q_{\bar a}(\x;m)\dots
B_{b_1}(\x,u_1)\prod_{j<m}\frac1{b(u_j-u_m)}\\
&&\times D_{b_m}(\x',u_m)\,K_{\bar ab_m}
\,R_{b_1b_m}(u_1-u_m)\dots R_{b_{m-1}b_m}(u_{m-1}-u_m)
\eea
With
$$
D_a(\x,u)\,g(\x,\uu)=\prod_{i\in I_+}b(x_i-u)\prod_{i\in I_-}
R_{ia}(x_i-u)\,g(\x,\uu),
$$
$$Q^{(1)}(\x^{(-)},\uu;b_m)={\rm tr}_a\,T^{Q(1)}_a(\x^{(-)},\uu;b_m)$$ (see
\ref{3.17a})), the assumption $(ii)^{(1)}$, in particular the relation
$$Q^{(1)}(\x^{(-)},\uu;b_m)\,f^{(1)}(\x^{(-)},\uu)=f^{(1)}(\x^{(-)},\uu')~,~~
(\uu'=u_1,\dots,u'_m=u_m+2)
$$
and the defining relations of the functions $\psi(x)$ and $\tau(x)$
(3.17) of [I] follows
$$
{\rm uw}_A^m(\x,\uu)\,g(\x,\uu)
=-{\rm tr}_{\bar a}~{\rm uw}_{D_{\bar a}}^m(\x,\uu')\,g(\x,\uu')
$$
which concludes the proof.

As in [I] we can iterate Theorem \ref{t3.1} to get the nested generalized Bethe
Ansatz with levels $k=1,\dots,N-1$.
To simplify the notation we introduce as an extension of
$I_\pm$ the index sets $I_k$ with $n_k=|I_k|$ elements for $k=0,\dots,N-1$
and as an extension of eqs.~(\ref{2.1}) and (\ref{3.15}) the spaces
\be{3.18}
{V^{(k)}}^{I_-I_k}=\bigotimes_{i\in I_-}\bar V^{(k)}_i\,
\bigotimes_{i\in I_k}V^{(k)}_i
\ee
with basis vectors $\mi\bar\alpha_1\dots\bar\alpha_{n_-}
\alpha_1\dots\alpha_{n_k}\ra~(k<\bar\alpha_i,\alpha_i\le N)$.
In terms of the previous notation we identify $I_+=I_0,~n_+=n_0,~
\V={V^{(0)}}^{I_-I_0}$ and $\VV={V^{(1)}}^{I_-I_1}$.
The Ansatz of level $k$ reads
\ban{3.19}
\lefteqn{{f^{(k-1)}}^{I_-I_{k-1}}
\left(\x^{(-)},\x^{(k-1)}\right)
=\sum_{\x^{(k)}}\,B^{(k-1)}_{I_-I_{k-1}\beta_{n_k}}
\left(\x^{(-)},\x^{(k-1)},x^{(k)}_{n_k}\right)\dots}\\
&&~~~~~\dots B^{(k-1)}_{I_-I_{k-1}\beta_1}
\left(\x^{(-)},\x^{(k-1)},x^{(k)}_1\right)\,
{g^{(k-1)}}^{I_-I_{k-1}\beta_1\dots\beta_{n_k}}
\left(\x^{(-)},\x^{(k-1)},\x^{(k)}\right)
\nn
\ean
where for $k<N-1$ analogously to Definition \ref{d3.2}
\ban{3.20}
{g^{(k-1)}}^{I_-I_{k-1}I_k}\left(\x^{(-)},\x^{(k-1)},\x^{(k)}\right)
=\prod_{i=1}^{n_{k-1}}\prod_{j=1}^{n_k}\psi(x_i^{(k-1)}-x_j^{(k)})~~~~~~~~~\\
\times\prod_{1\le i<j\le n_k}\tau(x^{(k)}_i-x^{(k)}_j)\,
{f^{(k)}}^{I_-I_k}(\x^{(-)},\x^{(k)})
\otimes{\Omega^{(k-1)}}^{I_{k-1}}.
\nn\ean
For $k=0$ with $f^{(0)}=f$ and $\x^{(0)}=\x^{(+)}$
we have to replace $\prod_{i=1}^{n_0}$ by  $\prod_{i\in I_+}$.
The start of the iteration is given by a $1\le k_{max}\le N$
such that all $n_k=0$ for $k\ge k_{max}$. 
We have to construct a vector valued function proportional to a fixed
vector which fulfills the assumptions of Theorem \ref{t3.1}.
This is given by 
\begin{lemm}\label{l3.1}
The vector valued functions for $k_{max}<N$
\be{3.21}
{f^{(k_{max}-1)}}^{I_-I_{k_{max}-1}}=\mi \bar N\dots
\bar Nk_{max}\dots k_{max}\ra
\ee
and for $k_{max}=N$
\be{3.22}
{f^{(N-1)}}^{I_-I_{N-1}}(\x^{(-)},\x^{(N-1)})
=\prod_{i\in I_-}\prod_{j\in I_{N-1}}\bar\psi_{N-1}(x^{(-)}_i-x_j^{(N-1)})\,
\mi \bar N\dots\bar NN\dots N\ra
\ee
fulfill the Conditions \ref{cond}.
The function $\bar\psi_{N-1}(x)$ has to obey the functional equation
\be{3.23}
(1+d(x))\,\bar\psi_{N-1}(x)=\bar\psi_{N-1}(x-2)
\ee
which is solved by
\be{3.24}
\bar\psi_{N-1}(x)=
\frac{\Gamma\left(\frac12+\frac x2\right)}
{\Gamma\left(\frac12+\frac1N+\frac x2\right)}.
\ee
Again the general solution is obtained by multiplication with an
arbitrary periodic function with period 2.
\end{lemm}
{\bf Proof:}
Conditions \ref{cond} (i) is fulfilled because of the symmetry with respect
to the particles and anti-particles among themselves.
Conditions \ref{cond} (ii) follows from the definition (\ref{3.3}) of the
$Q$-matrix and (\ref{2.4}) of the R-matrix.
For $k_{max}<N$ Conditions \ref{cond} (ii) follows since the Q-operators
act as the unit operator on the state ${f^{(k_{max}-1)}}^{I_-I_{k_{max}-1}}$.
For $k_{max}=N$ we have to take into account annihilation-creation
contributions of the anti-particle-particle R-matrix (\ref{2.4}).
For the last index in $I_-$ we have
$$
{Q^{(N-1)}}^{I_-I_{N-1}}(\x^{(-)},\x^{(N-1)};n_-)
=\prod_{j\in I_{N-1}}\,\Big(1+d(x^{(N-1)}_j-x^{(-)}_{n_-})\Big)\,\einsop
$$
and for the last index in $I_{N-1}$
$$
{Q^{(N-1)}}^{I_-I_{N-1}}(\x^{(-)},\x^{(N-1)};n_{N-1})
=\prod_{j\in I_-}\,\Big(1+d(x^{(-)}_j-x^{(N-1)}_{n_{N-1}}-2)\Big)\,\einsop
$$
which, together with eq.~(\ref{3.23}) implies the difference equations (ii)
for (\ref{3.22}).
%
\begin{coro}\label{c3.2}
The system of $U(N)$ matrix difference equations (\ref{3.2})
is solved by the generalized nested Bethe Ansatz (\ref{3.19}) with
(\ref{3.20}-\ref{3.23}).
\end{coro}

\setcounter{equation}{0}
\section{Weights of generalized $U(N)$ Bethe vectors}\label{s4}
The results of this section and also the techniques used are very similar
to the corresponding ones of [I]. Therefore we only point out the main
differences.
By an asymptotic expansion of the $R$-matrix and the monodromy
matrix $T$ (cf. eqs.(\ref{2.3}) and(\ref{2.9})) we get for $u\to\infty$
\ban{4.1}
R_{ab}(u)&=&\einsop_{ab}-\frac2{Nu}\,P_{ab}+ O(u^{-2})\\
T_{1\dots n,a}(\x,u)&=&\einsop_{1\dots n,a}+\frac2{Nu}\,M_{1\dots n,a}
+ O(u^{-2}).
\ean
Explicitly we get from eq.~(\ref{2.9})
\be{4.3}
M_{1\dots n,a}=\sum_{i\in I_+} P_{ia}-\sum_{i\in I_-} K_{ia},
\ee
where $I_\pm$ denote the particles and anti-particles, respectively.

In the following we will suppress the indices like $\s 1\dots n$.
In terms of matrix elements in the auxiliary space $V_a$ the generators
act on the basis states as
\be{4.4}
M^{\alpha'}_\alpha\mi\alpha_1,\dots,\alpha_i,\dots,\alpha_n\ra=
\left(\sum_{i\in I_+}\delta_{\alpha'\alpha_i}
-\sum_{i\in I_-}\delta_{\alpha'\alpha_i}\right)
\mi\alpha_1,\dots,\alpha,\dots,\alpha_n\ra.
\ee
The commutation relations of the $M$ and $T$ which follow
from the Yang-Baxter relations are the same as in [I].
Also Lemma 4.1 of [I] holds here for the $U(N)$ case. However, one has to
replace $\Omega\,g$ in eq.~(4.10) of [I] by $g$ as in BETHE ANSATZ \ref{be},
and the Q-matrix here is given by the trace of (\ref{3.17a}).

The slight difference to the $SU(N)$ case arises from the
action of the Cartan sub-algebra;  
i.e. from the diagonal elements of $M$ which are the weight operators
$W_\alpha=M^\alpha_\alpha$. They act on the basis vectors in $V$ as
\be{4.12}
W_\alpha\mi\alpha_1,\dots,\alpha_n\ra=
\left(\sum_{i\in I_+}\delta_{\alpha\alpha_i}
-\sum_{i\in I_-}\delta_{\alpha\alpha_i}\right)
\mi\alpha_1,\dots,\alpha_n\ra
\ee
which follows from eq.~(\ref{4.4}). In particular we have for the Bethe
Ansatz reference state $g$
\be{4.13}W_1\,g=n_+\,g.\ee
This means that also Lemma 4.2 of [I] holds here for the $U(N)$ case
where, however, in eq.~(4.16) of [I] the number $n$ is replaced by $n_+$. 
\begin{theo}
Let the vector valued function $f(\x)\in V$ be given by the BETHE ANSATZ
\ref{be} fulfilling the conditions of Theorem \ref{t3.1}, i.e. 
eqs.~(\ref{3.16}) and (\ref{3.17}).
If in addition $f^{(1)}$ is a highest weight vector and an eigenvector of
the weight operators with
\be{4.17}
W^{(1)}_\alpha\,f^{(1)}=w^{(1)}_\alpha\,f^{(1)},
\ee
then also $f$ is a highest weight vector
\be{4.14}
M^{\alpha'}_\alpha\,f=0~,~~(\alpha'>\alpha)
\ee
and an eigenvector of the weight operators
\be{4.15}
W_\alpha\,f=w_\alpha\,f~,~~~
w_\alpha=\cases{n_+-m &\rm for $\alpha=1$\cr
w^{(1)}_\alpha &\rm for $\alpha>1$}
\ee
with
\be{4.16}
w_\alpha\ge w_\beta~,~~~(1\le\alpha<\beta\le N),
\ee
\end{theo}
The proof of this theorem is again parallel to the corresponding one in [I].

The states $f^{(k_{max}-1)}$ of Lemma \ref{l3.1} which define the start
of the iteration of the nested Bethe Ansatz are obviously highest weight
states in $V^{(k_{max}-1)}$ with weight
$w^{(k_{max}-1)}_{k_{max}}=n_{k_{max}-1}-n_-$ by eq.~(\ref{4.12}).
\begin{coro}
If $f(\x)$ is a solution of
the system of $U(N)$ matrix difference equations (\ref{3.3})
$$
f(\dots,x_i+2,\dots)=Q(\x;i)\,f(\dots,x_i,\dots)~,~~(i=1,\dots,n)
$$
given by the generalized nested Bethe Ansatz of Corollary \ref{c3.2},
then $f$ is a highest weight vector with weights
\be{4.18}
w=(w_1,\dots,w_N)=(n_+-n_1,n_1-n_2,\dots,n_{N-2}-n_{N-1},n_{N-1}-n_-),
\ee
where $n_k$ is the number of $B^{(k)}$ operators in the Bethe Ansatz of level
$k$.
Further non-highest weight solutions of (\ref{3.2}) are given by
\be{4.20}
f^{\alpha'}_\alpha=M^{\alpha'}_\alpha\,f~,~~(\alpha'<\alpha).
\ee
\end{coro}
Note that in contrast to the $SU(N)$ case of [I], here the weights may
also be negative.

\setcounter{section}{4}\setcounter{equation}{0}
\section{Examples}\label{s5}
Similar as in [I] the solutions of the following examples may be multiplied
by a scalar function which is periodic in all $x$'s with period $2$ and
symmetric with respect to the anti-particle and particle variables.
\begin{exam}\label{e5.1}\rm
Let us consider the trivial case that there is no $B$-operator
in each level of the nested Bethe Ansatz, which means that $k_{max}$
of Section~\ref{s3} is equal to 1.
In the language of the conventional Bethe Ansatz for quantum chains
this correspond to the "ferro-magnetic vacuum". By Section
\ref{s4} this means that $\f(\x)$ has the weights
$$w=(n_+,0,\dots,0,-n_-).$$
For fixed positions of the particles ${I_+}$ and anti-particles ${I_-}$
by Lemma \ref{l3.1} the vector $\f\in V^{1\dots n}_{I_-}$ is given by
$$
f^{\alpha_1\dots\alpha_n}(\x)=\prod_{i\in I_-}\delta_{\alpha_i\bar N}\,
\prod_{i\in I_+}\delta_{\alpha_i1}.
$$
or if the anti-particles are sitting at the first places
$$
\f=\mi \bar N\dots\bar N1\dots 1\ra.
$$
This $\f$ is a highest weight vector in $V_{ref}$.
\end{exam}
\begin{exam}\label{e5.2}\rm
For $N>2$ and $n_+>2$ let us take the case where there is one $B$-operator
in the first level of the nested Bethe Ansatz and no $B$'s in higher levels,
which means that the $k_{max}$ of Section~\ref{s3} is equal to 2.
By Section~\ref{s4} this means that the weights are 
$$w=(n_+-1,1,0,\dots,0,-n_-).$$
The first level Ansatz reads
\be{5.1}
\f(\x)=\sum_u\,\B{\beta}(\x;u)~g^{1\dots n\beta}(\x;u).
\ee
For fixed positions of the particles ${I_+}$ and anti-particles ${I_-}$,
if the anti-particles are sitting on the left of all particles,
the function $g^{1\dots n\beta}$ is given by
\be{5.2}
g^{I_-I_+\beta}(\x,u)=\prod_{i\in I_+}\psi(x_i-u)\,
{f^{(1)}}^{I_-\beta}\otimes\Omega^{I_+}
\ee
with $\psi(x)=\frac{\Gamma(1-\frac1N+\frac x2)}{\Gamma(1+\frac x2)}$
(see eq.~(\ref{3.17}))
and by Lemma \ref{l3.1}
\be{5.3}
{f^{(1)}}^{I_-I_1}=\mi\bar N\dots\bar N2\ra~,~~~\Omega^{I_+}=\mi1\dots1\ra.
\ee
i.e. $f^{(1)}$ is the highest weight vector in $V^{(1)}_{ref}$.
Note that the function $\psi$ appears only with respect to the parameter
$x_i$ which correspond to the particles.
The action of the B-operators in eq.~(\ref{5.1}) can easily be obtained from
the definition of the R-matrices (\ref{2.3}) and (\ref{2.4}).
In particular we consider
\end{exam}
\begin{exam}\label{e5.3}\rm
As a simple case of Example \ref{e5.2} we take $n_-=1$ and $n_+=2$
$$
f^{\bar123}(x,y,z)=\sum_u\,\psi(y-u)\,\psi(z-u)\,\Big\{
c(y-u)\,b(z-u)\mi\bar N21\ra+c(z-u)\mi\bar N12\ra\Big\}.
$$
The sum over $u$ can be performed and gives the same result as in Example
5.3 of [I].
\end{exam}
\begin{exam}\label{e5.4}\rm
For $N=2$ let $n_-=n_+$ be $=1$.
In addition to the trivial case of Example \ref{e5.1} with no B-operator
there is only the possibility of Examples \ref{e5.2} and \ref{e5.3} with one
B-operator. By Section~\ref{s4} $f^{\bar12}(x,y)$ is an $U(2)$-singlet
vector with the weights $w=(0,0)$.
For $N=2$ we must take
into account the annihilation-creation contribution,
for the action of the B-operator on the reference state as well as
for $f^{(1)}$ due to Lemma \ref{l3.1}:
$$
f^{\bar12}(x,y)=\sum_u\,\bar\psi_1(x-u)\,\psi(y-u)\,\Big\{
d(x-u)\,b(y-u)\mi\bar11\ra+c(y-u)\mi\bar22\ra\Big\}
$$
with $\psi(x)=\bar\psi_1(x)=\frac{\Gamma(\frac12+\frac x2)}
{\Gamma(1+\frac x2)}$.
Similar as in Example \ref{e5.3} the sum over $u=\tilde u-2l,~(l\in{\bf Z})$
can be performed
$$
f^{\bar12}(x,y)=\frac{\cos\frac\pi2(x-y)}
{\cos\frac\pi2(x-\tilde u)\cos\frac\pi2(y-\tilde u)}\,\frac1{x-y+1}
\,\Big\{\mi\bar11\ra+\mi\bar22\ra\Big\}.
$$
As a generalization of this formula for arbitrary $N$ we consider
\end{exam}
\begin{exam}\label{e5.5}\rm
Let us take for  $n_-=n_+=1$ the case where there is exactly one $B$-operator
in each level of the nested Bethe Ansatz,
which means that the $k_{max}$ of Section~\ref{s3} is equal to $N-1$.
By Section~\ref{s4} this means the weights are $w=(0,\dots,0)$, i.e.
$f^{12}(x,y)$ is an $U(N)$-singlet.
The first level Ansatz reads
\be{5.4}
f^{\bar12}(x,y)=\sum_u\,B_{\bar12,\beta}(x,y;u)~g^{\bar12\beta}(x,y;u)
\ee
where $u=\tilde u+2l$, $l\in{\bf Z}$ and
\be{5.5}
g^{\bar120}(x,y,u)=\psi(y-u)\,{f^{(1)}}^{\bar10}(x,u)\,\Omega^2.
\ee
The higher level Ans\"atze are of the same form. One solution is
\be{5.6}
f^{\bar\alpha\beta}(x,y)=\frac1{x-y+1}\,\delta_{\alpha\beta}.
\ee
\end{exam}
{\bf Proof:}
The solution (\ref{5.6}) is a up to a constant the particular case $k=0$ of
a general formula valid for all levels.
This general formula (again up to unimportant constants)
\be{5.7}
{f^{(k)}}^{\bar\alpha\beta}(x,y)=\bar\psi_k(x-y)\,\delta_{\alpha\beta}
~,~~(k<\alpha,\beta\le N)
\ee
where as an extension of eq.~(\ref{3.24})
\be{5.8}
\bar\psi_k(x)=
\frac{\Gamma\left(\frac12+\frac x2\right)}
{\Gamma\left(\frac32-\frac kN+\frac x2\right)}
\ee
will be proved inductively. For $k=N-1$ as the start of the iteration formula
(\ref{5.7}) follows from Lemma \ref{l3.1}. For the other values of $k$ it
follows recursively from eqs.~(\ref{3.19}) and (\ref{3.20}):
\ban{5.9}
{f^{(k-1)}}^{\bar12}(x,y)&=&\sum_uB^{(k-1)}_{\bar12\beta}(x,y;u)\,
{g^{(k-1)}}^{\bar12\beta}(x,y;u)\\\label{5.10}
\sum_uB^{(k-1)}_{\bar12\beta}(x,y;u)\mi\bar\alpha k\ra
&=&d(x-u)\,b(y-u)\mi\bar kk\ra+c(y-u)\mi\bar\alpha\beta\ra\\\label{5.11}
{g^{(k-1)}}^{\bar\alpha\beta\gamma}(x,y;u)
&=&\psi(y-u)\,{f^{(k)}}^{\bar\alpha\gamma}(x,u)\,\delta_{\beta2}.
\ean
We calculate the right hand side of eq.~(\ref{5.9}) inserting
eqs.~(\ref{5.10})
and (\ref{5.11}) with ${f^{(k)}}$ given by (\ref{5.7}) and with $u=x+2l$
\eqa{5.12}{
\lefteqn{
\sum_l\psi(x-u)\,\bar\psi_k(y-x)\,
\Big\{d(x-u)\,b(y-u)\,(N-k)\mi\bar kk\ra
+c(y-u)\,\sum_{\alpha=k+1}^N\mi\bar\alpha\alpha\ra\Big\}}\\
&=&\frac1N\sum_l\left\{
\frac{\Gamma\left(-\frac1N+l\right)}{\Gamma\left(l\right)}
\frac{\Gamma\left(-\frac12+\frac{x-y}2+l\right)}
{\Gamma\left(\frac32-\frac kN+\frac{x-y}2+l\right)}
\,(N-k)\mi\bar kk\ra\right.\\
&&~~~~~~~~+\left.
\frac{\Gamma\left(-\frac1N+l\right)}{\Gamma\left(1+l\right)}
\frac{\Gamma\left(\frac12+\frac{x-y}2+l\right)}
{\Gamma\left(\frac32-\frac kN+\frac{x-y}2+l\right)}
\,\sum_{\alpha=k+1}^N\mi\bar\alpha\alpha\ra\right\}~~~~~~~~~~~~~~~~~\\
&=&const.\,\bar\psi_{k-1}(x-y)\,\sum_{\alpha=k}^N\mi\bar\alpha\alpha\ra
=const.\,{f^{(k-1)}}^{\bar12}(x,y)
}
The sums over $l$ have been performed using the Gauss formula
$$
\sum_l\frac{\Gamma(a+l)\Gamma(b+l)}{\Gamma(l)\Gamma(c+l)}=
\frac{\Gamma(c-a-b-1)\Gamma(1+a)\Gamma(1+b)}{\Gamma(c-a)\Gamma(c-b)}
$$
This concludes the proof of formula \ref{5.7}.
\begin{exam}\label{e5.6}\rm The last example will be used in \cite{BFK} to
calculate the exact three particle form factor of the fundamental field
in the chiral Gross-Neveu model.
We take $n_-=1,~n_+=2$ and again exactly one $B$-operator in each level
of the nested Bethe Ansatz.
By Section~\ref{s4} this means that $f^{\bar123}(x,y,z)$ is an $U(N)$-vector
with weights
$$w=(1,0,\dots,0).$$
The first level Ansatz reads
\be{5.13}
f^{\bar123}(x,y,z)=\sum_u\,B_{\bar123,\beta}(x,y,z;u)
\,g^{\bar123\beta}(x,y,z;u)
\ee
The higher level $(k>1)$ Bethe Ansatz coincides with that of
Example \ref{e5.5}.
A solution of the difference equations for this case is given by
\ban{5.14}\nn
f^{\bar123}(x,y,z)&=&\sum_u\left\{
\frac{\Gamma\left(-\frac1N+\frac{x-u}2\right)}
{\Gamma\left(\frac{x-u}2\right)}
\frac{\Gamma\left(-\frac1N+\frac{y-u}2\right)}
{\Gamma\left(\frac{y-u}2\right)}
\frac{\Gamma\left(-\frac12+\frac{z-u}2\right)}
{\Gamma\left(\frac32-\frac1N+\frac{z-u}2\right)}
\,(N-1)\,\mi\bar111\ra~~~~~\right.\\
&&-~\frac{\Gamma\left(-\frac1N+\frac{x-u}2\right)}
{\Gamma\left(\frac{x-u}2\right)}
\frac{\Gamma\left(-\frac1N+\frac{y-u}2\right)}
{\Gamma\left(1+\frac{y-u}2\right)}
\frac{\Gamma\left(\frac12+\frac{z-u}2\right)}
{\Gamma\left(\frac32-\frac1N+\frac{z-u}2\right)}
\,\sum_{\alpha=2}^N
\mi\bar\alpha\alpha1\ra\\
&&-~\left.\frac{\Gamma\left(-\frac1N+\frac{x-u}2\right)}
{\Gamma\left(1+\frac{x-u}2\right)}
\frac{\Gamma\left(1-\frac1N+\frac{y-u}2\right)}
{\Gamma\left(1+\frac{y-u}2\right)}
\frac{\Gamma\left(\frac12+\frac{z-u}2\right)}
{\Gamma\left(\frac32-\frac1N+\frac{z-u}2\right)}
\,\sum_{\alpha=2}^N\mi\bar\alpha1\alpha\ra\right\}\nonumber
\ean
where $u=\tilde u+2l$, $l\in{\bf Z}$.
\end{exam}
{\bf Proof:}
Analogously to eqs.~(\ref{5.10}) and (\ref{5.11}) we have for $(k=1)$
\bea
B_{\bar123,\beta}(x,y,z;u)\mi\bar\alpha11\ra
&=&b(x-u)\,d(y-u)\,b(z-u)\,\delta_{\alpha\beta}\mi\bar111\ra\\
&&+~c(y-u)\,b(z-u)\mi\bar\alpha\beta1\ra\\
&&+~c(z-u)\mi\bar\alpha1\beta\ra
\eea
$$
g^{\bar\alpha\beta}(x,y,z;u)
=\psi(y-u)\psi(z-u)\,{f^{(1)}}^{\bar\alpha\beta}(x,u)
$$
where  ${f^{(1)}}^{\bar\alpha\beta}(x,u)$ is the same function as in
Example \ref{e5.5} and given by eq.~(\ref{5.7}) for $k=1$.
Inserting this into eq.~(\ref{5.13}) we get
in analogy to eq.~(\ref{5.12}) the result (\ref{5.14}).
\\[1cm]
{\bf Acknowledgment:} The authors have profited from discussions with
A.~Fring, R.~Schra\-der, F.~Smirnov and A.~Belavin.


\begin{thebibliography}{9999}

\bibitem
{BKZ} H.M. Babujian, M. Karowski and Zapletal,
'$SU(N)$ Matrix Difference Equations and a Nested Bethe Ansatz',
to be published.


\bibitem
{BFK} H.M. Babujian, A. Fring and M. Karowski, 'Form factors of the
$SU(N)$ - chiral Gross-Neveu model', in preparation.

\bibitem
{BKKW}B. Berg, M. Karowski, V. Kurak and P. Weisz,
Nucl. Phys. {\bf B134} (1978) 125-132.

\bibitem
{KW} M. Karowski and P.Weisz,
Nucl. Phys. {\bf B139} (1978) 445-476.



\end{thebibliography}
\end{document}